# NEW EVIDENCE FOR THE COSMOLOGICAL ORIGIN OF γ-RAY BURSTS

Tsafrir Kolatt[1] and Tsvi Piran[2]

[1] Harvard-Smithsonian Center for Astrophysics, 60 Garden St. Cambridge, MA 02138 USA

[2] Racah Institute for Physics, The Hebrew University, Jerusalem, 91904 Israel

## Abstract

We find that Gamma-ray bursts (GRBs) at the 3B catalog are correlated (at 95% confidence level) with Abell clusters. This is the first known correlation of GRBs with any other astronomical population. It confirms the cosmological origin of GRBs. Comparison of the rich clusters auto-correlation with the cross-correlation found here suggests that ∼ **26 ± 15 %** of an accurate ($\delta < 2.3°$) position GRBs sub-sample members are located within **600 h$^{-1}$Mpc.**

## 1 Introduction

One of the major obstacles in GRBs research is the lack of any association of GRBs with any other astronomical population. The recent observations of the BATSE experiment on the COMPTON-GRO observatory suggest that GRBs are cosmological (Meegan et al. 1992.) The GRBs distribution appears to be isotropic and there is a paucity of weak bursts. Both facts are naturally explained by a cosmological distribution. The observed peak flux distribution in the BATSE catalog agrees well with a theoretical cosmological peak flux distribution (Piran 1992; Mao & Paczyński 1992; Dermer 1992; Wickramasinghe et al. 1993; Cohen & Piran 1995.) These facts can, however, be accommodated by some extended

Galactic halo models and there is an ongoing debate on the galactic or extra-galactic origin of GRBs (Paczyński 1995; Lamb 1995.)

We have discovered a positive cross-correlation signal between the GRBs distribution in the 3B catalog and Abell clusters. This confirms our previous result that indicated the existence of such a correlation in the BATSE 2B catalog (Cohen, Kolatt, & Piran 1994.) This correlation demonstrates the cosmological origin of GRBs. It also enables us to estimate directly the distance scale to GRBs. We did not find any significant auto-correlation of GRBs in the 3B catalog. Since GRBs are distributed over cosmological distances ($z \approx 1$) one should not expect a cross-correlation with galaxy catalogs that consist of relatively nearby objects. Indeed we did not find any cross-correlation with IRAS galaxies neither with galaxies in any other optical catalog. However, The volume limited sample of Abell clusters (Abell 1958; Abell, Corwin, & Olwin 1989) extends to larger distances and may lead to a detectable cross-correlation.

## 2 The Data

The data consist of 3616 Abell clusters of richness class $R \geq 0$ in Galactic latitudes $|b| > 30°$. The cutoff is meant to reduce statistical noise emerging from low values of the cluster selection function $\phi_{cl}$ (Scaramella *et al.* 1991.) The clusters show a strong auto-correlation signal (Bahcall & Soneira 1983; Batuski *et al.* 1989; Peacock & West 1992), and their typical redshift values are $z \approx 0.15 - 0.2$ . There are 549 GRBs in the 3B catalog with $|b| > 30°$. The GRBs selection function, $\phi_{grb}$, is anisotropic due to different exposure durations in different directions. It can be derived from the exposure duration ratio as function of declination and has up to 20% variation across the sky. The position of the bursts is uncertain, with an angular error for the $j^{th}$ burst given by $\delta_j = \sqrt{\delta_{sys}^2 + \delta_{stat,j}^2}$ , where $\delta_{sys} = 1.6°$ is a systematic error and $\delta_{stat,j}$ is a statistical error that varies from one burst to another. The statistical error, $\delta_{stat,j}$, decreases with the burst strength. For a weak burst $\delta_{stat,j}$ can be quite large (up to 20°) and it might be meaningless to include objects with such poorly known positions in our analysis. Hence we have constructed an "accurate" sub-class of the GRBs sample, that contains 136 GRBs with a good positional accuracy, namely for which the positional error $\delta_j$ satisfies: $\delta_j \leq \delta_{max} \equiv \sqrt{2}\delta_{sys}$. Clearly, because of the nature of this sub-class it contains stronger bursts on average.

## 3 Cross Correlation Estimate

The cross-correlation of the Abell clusters and the anisotropic sky coverage of the BATSE data make it difficult to estimate the cross-correlation between these two data sets in the usual way. We do like however to start up with the conventional cross-correlation evaluation in order to obtain its model independent values. Let $N_{grb-cl}(\theta)$ be the number of pairs of GRBs and clusters separated by the angle $\theta$ where for the $i^{th}$ bin $\theta_{i-1} < \theta < \theta_i$ and let $N_{grb-po}(\theta)$ be the number of pairs of GRBs and randomly distributed particles subject to the cluster selection function $\phi_{cl}$, then the angular cross-correlation function, $w(\theta)$, is defined as

$$1 + w(\theta) = \frac{N_{grb-cl}(\theta)}{N_{grb-po}(\theta)} \frac{n_{po}}{n_{cl}}, \qquad (1)$$

where $n_{cl}$ and $n_{po}$ are the number densities of the clusters and Poisson particles respectively. In order to obtain a bias free estimate of the correlation, we corrected for the GRBs selection function by assigning a weight to each pair count inversely proportional to $\phi_{grb}^j$.

The main advantage of the cross-correlation assessment is that it allows to reduce the noise level appreciably in comparison to the GRBs auto-correlation. In the limit where $n_{po} \gg n_{cl}$, the Poisson error in $w(\theta)$ is $(\Delta w)_p = (1 + w)(N_{grb-cl})^{-1/2}$ for each $\theta$ bin. The overall error in the correlation value is larger than the Poisson error due to the cluster auto-correlation, the possible GRBs auto-correlation, and the assigned weights. Considering the cluster auto-correlation alone, we get a rough estimate for the error multiplicative factor (*e.g.*Olivier *et al.* 1990) by letting $\Sigma$ be the cluster mean surface density and $w_{cl-cl}(\theta) \simeq A\theta^{-1}$. We then evaluate

$$u \simeq \Sigma \int d\phi \int_{\theta_{i-1}}^{\theta_i} w_{cl-cl}(\theta) d(\cos\theta), \qquad (2)$$

to get $\Delta w \simeq (1+u)^{1/2}(\Delta w)_p$. For $\theta$ expressed in degrees, $A \approx 0.7$ (*cf.* Batuski *et al.* 1989 for $R \geq 1$ and here $R \geq 0$) and a single 4° radius bin, we obtain $\sqrt{1+u} \simeq 2.0$. Though this estimate is only approximate, it allows a model-free interpretation of our results later on.

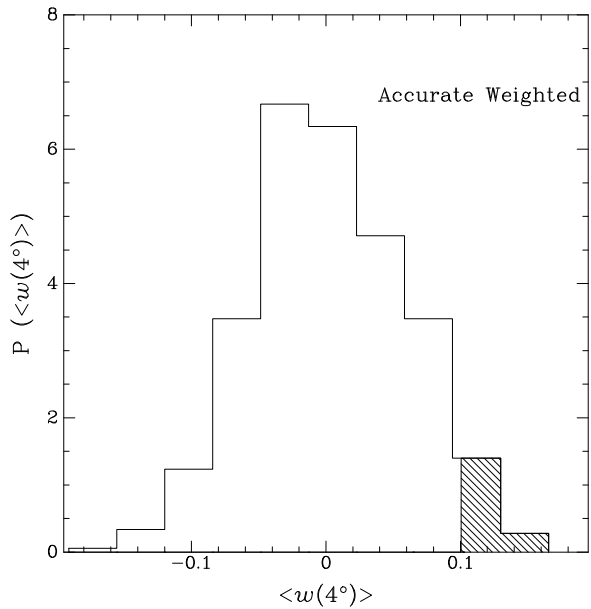

Figure 1: The probability distribution for random artificial GRBs sample and Abell clusters cross-correlation. Only accurate GRBs weighted by their positional errors were used for the evaluation. The shaded area shows the 5% probability for random GRBs sample to produce higher cross-correlation signal than the true GRBs sample signal.

In order to circumvent the difficulties in the cross-correlation evaluation, we have instead tested the null hypothesis, namely that the GRBs show no correlation with the rich clusters. We calculate $N_{grb-cl}(\theta)$ the number of GRBs−cluster pairs whose separation is smaller than a given angle $\theta$ for $\theta = 1°, 2°, ..., 6°$. We then create 500 random Poissonian realizations (artificial catalogs = "ac") of particles with the same selection function and error distribution as the GRBs and calculate $N^i_{ac-cl}(\theta)$ for each realization $i$. We then attempt to rule out the null hypothesis by considering the fraction of random artificial catalogs for which $N_{grb-cl}(\theta) > N^i_{ac-cl}(\theta)$. This provides us with a direct estimate of the null hypothesis rejection level.

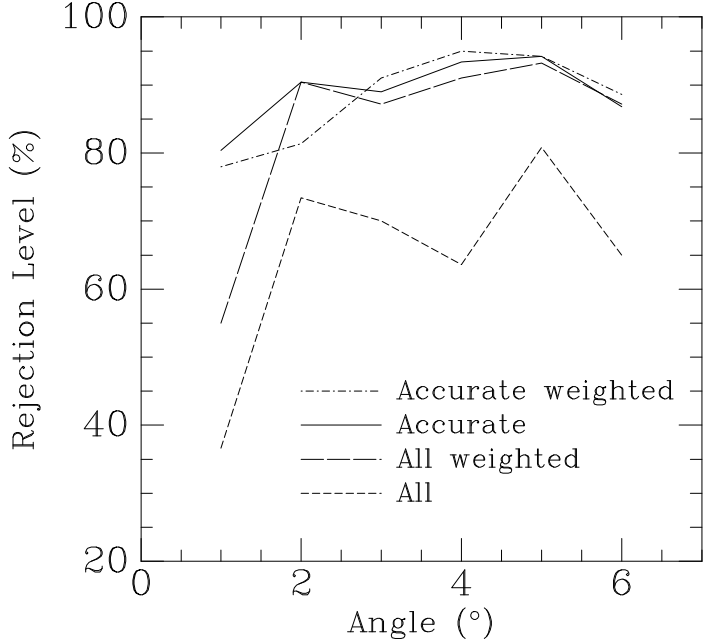

Figure 2: Rejection levels for the null hypothesis of no cross-correlation between GRBs and rich clusters. Different lines are for all GRBs and a sub-class of accurate position GRBs weighted or not by their positional errors.

Since we conduct a purely statistical experiment we try to maximize the statistical significance by performing the same analysis for the full GRBs sample ($|b| > 30°$) and for the "accurate" sub-class. In each of the above cases we have also carried out calculation for weighted correlation weighting each GRB - cluster pair by the GRB error, $\delta_j^{-2}$. Figure 1 shows the entire probability distribution for cross-correlation between artificial catalogs of the GRBs accurate weighted sample and the Abell clusters catalog. Only in 5% out of the 500 artificial catalogs a correlation signal higher than the true correlation was found. The results for the different sub-samples and weighting schemes are summarized in table 1 and plotted in figure 2.

Checking for self consistency within the table entries, we notice that for the last case ("accurate" sub-class with statistical weighted cross-correlation) where we find the largest statistical significance, the maximum is obtained for 4° top-hat on the sky. The same holds for the second case ("accurate" sub-class with a regular correlation function) and for the third case (all bursts with weighted correlation function). These results reflect the combined effect of cross-correlation accompanied by systematic and random errors. Even if all GRBs reside in Abell clusters we could not expect higher correlation at smaller angles because of the errors. The statistical significance increases when we use weighted correlation function (column 5 vs. column 3) and when we use the accurate sub-class in place of the whole sample (column 5 vs. column 4 and column 3 vs. column 2). Both trends are reassuring. We recall that the more accurate bursts are stronger on average, hence nearer and therefore more correlated with the clusters.

Table 1: Significance of cross-correlation coefficient (%)

|  | All regular | "Accurate" regular | All weighted | "Accurate" weighted |
|---|---|---|---|---|
| number | 549 | 136 | 549 | 136 |
| 1° | 36.6 | 80.4 | 55.0 | 78.0 |
| 2° | 73.4 | 90.4 | 90.4 | 81.4 |
| 3° | 70.0 | 89.0 | 87.2 | 91.0 |
| 4° | 63.6 | 93.4 | 91.0 | 95.0 |
| 5° | 80.8 | 94.2 | 93.2 | 94.2 |
| 6° | 65.0 | 86.8 | 87.2 | 88.6 |

## 4  The Inferred Distances

Having found a cross-correlation signal and the optimal angle to look for it, we can estimate the GRBs fraction that actually contribute to the signal, *i.e.* overlaps with the cluster sample volume. In a model where all GRBs reside in clusters, we compare the ratio between the detected cross-correlation and the two-dimensional top-hat average over the cluster auto-correlation. For a circle of radius $\theta_c$ we get $\langle w_{cl-cl}(\theta_c) \rangle \simeq 2A\theta_c^{-1}$ and $\langle w_{cl-cl}(4°) \rangle \simeq 0.35$. The measured, minimally biased values for the cross-correlation at a 4° bin are 0.013 and 0.093 for all GRBs and for the "accurate" sub-sample respectively. While the former is not much of a use due to signal-to-noise ratio of $\sim 1$, the error estimate for the latter is $\Delta w(4°) \simeq 0.053$. Notice that the statistical significance of the measurement has already been established, based on the null hypothesis rejection, and the current analysis is needed for the overlap estimate only. The overlapping fraction is hence $\sim 0.093/0.35 = 0.26 \pm 0.15$. For the "accurate" sub-sample with 136 GRBs, we conclude that $35 \pm 20$ of them lie in a volume of a 600 $h^{-1}$Mpc sphere. Using the relative amplitudes, and the expression for the expected angular cross-correlation (Lahav, Nemiroff, & Piran 1990, equation 9) one gets a characteristic depth for the GRBs "accurate" sub-sample of $R_{*,grb-a} = 940^{+300}_{-120} h^{-1}$Mpc *i.e.* $z_{grb-a} = 0.31^{+0.1}_{-0.04}$. In a different estimates we calculate $n(z)$, the fraction of observed bursts up to a redshift $z$ (*e.g.* Cohen & Piran 1995) in a given cosmological model. Given the fact there are $35 \pm 20$ GRBs within $z = 0.2$, we obtain for $\Omega = 1$ and no source evolution $z_{max-a} = 0.36^{+0.27}_{-0.07}$ in which BATSE will detect 156 bursts. The equivalent estimate for all GRBs yields $\sim 4\%$ overlap with the cluster sample, namely $\sim 22$ GRBs within the same volume but with a large error. The much larger error can be interpreted as a steep build-up of the noise level due to non-overlapping GRBs. For the whole sample this implies a maximum detection redshift of $z_{max} = 0.7^{+0.93}_{-0.07}$. The two estimates are consistent with each other and suggest that the main correlation contribution for the entire GRBs sample essentially comes from the "accurate" sub-sample. An equivalent calculation, using the cluster-galaxy correlation function (Mo, Peacock, & Xia 1993; Lilje, & Efstathiou 1988), with approximately half the cluster auto-correlation function amplitude in the relevant range and under the assumption that all GRBs reside in galaxies, leads to a larger overlapping fraction for the "accurate" sub-sample of $\sim 0.5$ and to lower $z$ values.

## 5 Conclusions and Discussion

We have found a cross-correlation of GRBs with Abell clusters. The large rich cluster sample helps to overcome the statistical noise that prevents us from finding any GRBs auto-correlation. The correlation we found is believed to be a true spatial correlation. However, weak lensing magnification bias could also contribute to the observed effect. The predicted magnification correlation averaged on a $\theta_c$ radius window is given by (*e.g.* Bartelmann 1995),

$$\langle w(\theta_c) \rangle = 2\theta_c^{-2}(-a-1)b \int_0^{\theta_c} w_{\mu\delta}(\theta) d(\cos\theta), \qquad (3)$$

where $w_{\mu\delta}$ is the matter-magnification correlation, $b$ the considered lenses biasing factor and $a$ is the power law index for the intrinsic GRBs source counts as a function of their flux. For the most favorable case for magnification we assume all GRBs are at $z = 1.5$, consider $z < 0.2$ clusters along the line of sight with a biasing factor $b = 5$, and use the standard CDM model function for these parameters (Bartelmann 1995) of $w_{\mu\delta} \simeq 7 \times 10^{-4} \theta^{-0.6}$. The obtained value is then $\langle w(4°) \rangle \leq (-a-1)0.014$, which is one sixth of the detected signal for the accurate sample, and comparable to the (uncertain) signal for all GRBs (for reasonable values of $a$).

If real spatial correlation is responsible for the detected signal, we know that Abell clusters extend up to $z \approx 0.15 - 0.2$ and therefore the relative amplitudes of the cluster auto-correlation and the GRBs cluster cross-correlation provides an estimate for the GRBs redshift distribution. For a sub-sample of accurately measured GRBs we found that $z_{max-a}$ of this population is $0.36^{+0.27}_{-0.07}$ if all accurate GRBs are in clusters and $z_{max-a} \simeq 0.27^{+0.12}_{-0.05}$ if they are all in galaxies. For the whole GRBs sample this implies $z_{max} = 0.7^{+0.93}_{-0.07}$ if GRBs originate in clusters and $0.5^{+0.3}_{-0.15}$ if they are all in galaxies. The upper limits of these values are within the range of $z_{max}$ estimates as obtained from fitting the peak-flux distribution to a cosmological distribution. Cohen and Piran (1995) find $z_{max} = 2.1^{+1.0}_{-0.7}$ for the long bursts and $z_{max} \leq 0.5$ for the short bursts, while Fenimore *et al.* (1993) find $z_{max} \approx 0.9$ for the combined population. However, In a more recent analysis Fenimore & Bloom (1995) find that if GRB time dilation is entirely attributed to cosmological redshift streching, then $z_{max} > 6$. Here we have a combined population of long and short bursts. The average values are on the low side and this may indicate that lensing magnification and not just spatial correlation contributes to the observed signal. More data is needed to decrease the error estimates and to determine which of the two possibilities is the dominant contribution.

The cross-correlation is the first association of GRBs with any other population of astronomical objects. It demonstrates, of course, the cosmological origin of the bursts. It does not mean, however, that the GRBs originate necessarily in clusters, as it is consistent with the picture in which the bursts originate in galaxies or galaxy halos that in turn, are strongly correlated with Abell clusters.

We thank Ehud Cohen, Ramesh Narayan, Reem Sari, and Eli Waxman for helpful discussions. This research was supported in part by the US National Science Foundation (PHY 91-06678) and by the Israeli National Science Foundation.